# Quantification of Probe-Sample Electrostatic Forces with Dynamic Atomic Force Microscopy

Nina Balke,[1*] Stephen Jesse,[1] Ben Carmichael,[2] M. Baris Okatan,[1] Ivan I. Kravchenko,[1] Sergei V. Kalinin,[1] Alexander Tselev[1,3]

[1]Center for Nanophase Materials Sciences, Oak Ridge National Laboratory, Oak Ridge, TN 37831 United States

[2]Southern Research Institute, Birmingham, AL 35211, United States

[3]CICECO and Department of Physics, University of Aveiro, Aveiro, 3810-193, Portugal

*Corresponding author: balken@ornl.gov

## Abstract

Atomic Force Microscopy (AFM) methods utilizing resonant mechanical vibrations of cantilevers in contact with a sample surface have shown sensitivities as high as few picometers for detecting surface displacements. Such a high sensitivity is harnessed in several AFM imaging modes. Here, we demonstrate a cantilever-resonance-based method to quantify electrostatic forces on a probe arising in the presence of a surface potential or when a bias voltage is applied to the AFM probe. We find that the electrostatic forces acting on the probe tip apex can strongly dominate cantilever response in electromechanical measurements and produce signals equivalent to few pm of surface displacement. In combination with modeling, the measurements of the force were used to determine the strength of the electrical field at the probe tip apex in contact with a sample. We find an evidence that the electric field strength in the junctions is limited by about 0.5 V/nm. This field can be sufficiently strong to significantly influence material states and kinetic processes through charge injection, Maxwell stress, shifts of phase equilibria, and reduction of energy barriers for activated processes. Besides, the results provide a baseline for accounting for the effects of local electrostatic forces in electromechanical AFM measurements as well as offer additional means to probe ionic mobility and field-induced phenomena in solids.

**Introduction**

In a number of AFM imaging modalities, nanometer-scale sample properties are probed locally with a conducting probe in contact with a sample when the probe is under a voltage bias in respect to another electrode. In particular, in electromechanical measurements, surface displacements induced by a localized ac electric field at the probe tip are detected as a function of a dc voltage applied to the probe [1-8]. A similar setup is used for nanometer-scale material modifications [9-11]. Concentrated electric fields developed during voltage application in the tip-sample junction in a volume of a few tens of nanometers in size can be extremely strong – in excess of $10^8$ V/m, which has been utilized to explore phenomena and material behaviors in strong electric fields, for example, nanoscale melting [12], electric-field-induced water condensation [13, 14], water dissociation [15], nanoscale lubrication [16], and ionic conduction [17]. Such studies would highly benefit from knowledge of the electric field strength at the tip-sample contact through providing well-controlled stimulus as well as for correct interpretation and modeling of results. Observation of effects associated with the large electric field in the tip-sample junction indirectly proves the large strength of the field (see, e.g., refs. [9, 13, 14, 16]). However, currently there are no methods, which would allow measuring the electric field strength in the junction. Such a method could be of use for estimating the upper limits of the field developed in the junction, the electrostatic force acting on the tip apex at the contact, threshold field values necessary for material modifications, and other tasks.

On the other hand, many dynamic scanning probe measurements rely on resonance enhancement of measurement sensitivity when measurements are carried out with use of a harmonic stimulus with a frequency near one of the vibrational resonant modes of the cantilever [18-20]. The resonant enhancement makes the measurements highly sensitive—tip displacements in sub-pm range can be detected—and special attention should be given to possible parasitic effects, which may distort the measurement results [21-30]. In the electromechanical measurements, one of such effects is the presence of the electrostatic force acting on all parts of the cantilever probe: the lever, tip, and tip apex. The electric field developing under a tip apex can result in a force large enough to produce detectable apex displacements against the mechanical stiffness of the tip-sample contact [31], especially in experiments with a strongly polarizable material, such as a high-permittivity dielectric [32]. The signal associated with this effect is not easy to distinguish from the signal due to the sample surface displacements. It is worth noting that historically, this fact led to two alternative interpretations and a debate on the signal origin in the piezoresponse force microscopy of ferroelectric materials [2, 31, 33, 34]. The electrostatic force on the tip apex is dependent on the local dielectric properties—permittivity and conductivity—as well as surface topography and can produce local contrasts superimposed onto the electromechanical response. Furthermore, the electric field in the junction in excess of $10^7$ V/m leads to charge injection into the sample. The injected charge alters the sample surface potential and imposes a localized electrostatic force on the tip even without application of a voltage bias. Variability of the injected charge behavior across the sample surface, differences in charge dissipation and retention between materials, surface conditions, and sample features may appear in the images as well as a contrast superimposed onto the electromechanical response. The strength of these effects may overshadow responses of interest such as those due to the converse piezoelectric effect (in piezoresponse force microscopy [31, 34-36]), electrostriction, or Vegard expansion (in electrochemical strain microscopy [7, 37-40]). Obviously, quantification of the electrostatic force may help in managing and accounting for the associated side effects.

In this paper, we introduce and implement a method for quantification of the electrostatic force acting on the tip apex. An AFM probe is used as a dynamic dynamometer. The method employs measurements of the tip apex displacements induced by a harmonic electrostatic force with simultaneous determination of the tip-sample contact stiffness. As a part of the described implementation, an electrostatic force on the tip apex is created by injecting charges into a dielectric film by the same tip shortly before the force measurements. This enables measurements of the force without application of a dc bias to the cantilever probe, which insures localization of the electrostatic force on the tip apex. Further, combining the measurements with the modeling allows estimation of the electric field strength in the voltage-biased tip-sample junction based on the measured value of the electrostatic force.

The described method employs cantilever resonances at the first flexural contact resonant mode. Similar to a previous publication on quantification of surface displacements driving the cantilever vibrations in a similar sample-probe configuration [41], here, quantification of the tip apex displacements was performed via static calibration of cantilever sensitivity and application of a factor to account for the shape of the resonant mode in the dynamic measurements at a vibrational resonance (which depends on properties of cantilever and tip-sample contact). The tip-sample contact stiffness was determined through the measured resonant frequency of the vibrational mode. The measured values of the tip apex displacements and contact stiffness are used to calculate the electrostatic force at the tip apex. The experiments were performed on an amorphous $HfO_2$ thin films sputtered on a $Au/SiO_2/Si$ substrate. In the experiments, we find that in the contact-mode measurements on a thin film with a bottom electrode, only electrostatic forces acting on the tip apex have to be taken into account, with a negligible contribution from the cantilever shank. We find an evidence that the electric filed strength in the junctions is limited by about 0.5 V/nm.

**Measurements of tip displacements induced by electrostatic force**

In the analysis and measurements of the electrostatic forces acting on a cantilever AFM probe in contact with a sample, we adapt the description of the electrostatic forces developed for traditional non-contact AFM measurements such as electrostatic force microscopy (EFM) or Kelvin probe force microscopy (KPFM). Since the Coulombic force is conservative, the electrostatic force acting on the probe $\mathbf{F}_{es}$ is expressed as a negative gradient of the electrostatic energy, $W_{es}$, over the relative positions of the tip and a counter electrode:

$$\mathbf{F}_{es} = -\nabla W_{es} = -\frac{1}{2}\nabla(CV^2) = -\frac{1}{2}\nabla C \cdot V^2$$

where $\nabla$ is the nabla operator, $C$ is a probe capacitance in respect to the counter-electrode, and $V$ is a bias voltage applied to the probe vs. the counter electrode. In EFM and KPFM, the gradient of $C$ is taken equal to the derivative $dC/dz$ of the capacitance as a function of the probe-sample distance $z \neq 0$. However, when the probe tip is in contact with a sample, and $z = 0$, the derivative d$C$/d$z$ is undefined. This is easy to see by making an attempt to construct the derivative with a virtual displacement δ$z$ without violating the constraint of the mechanical contact between the tip and sample. In an idealized case of hard, non-deformable tip and sample, such a displacement assumes penetration of the tip into the sample, and while the energy $W_{es}$ is a continuous function of the tip position, its gradient $\nabla W_{es}$ will experience a jump at the tip-sample contact, analogously to the electric field **E** at a dielectric-vacuum interface. In reality, both the tip and sample will be deformed after they are brought in contact. This effectively removes the jump in $\nabla W_{es}$, however, the details of the deformation will be dependent (consistently with the force) on mechanical properties of the tip and sample materials, contact shape, as well as presence of adhesive forces in the junction, the effective sharpness of the interface, and other factors. Given these details cannot be determined *a priory*, we conclude, that in experiments with a tip in contact with a sample, $dC/dz \equiv C'$ has to be viewed as a proportionality coefficient between the force $F_{es}$ and voltage $V$ in $F_{es} = \frac{1}{2} C' \cdot V^2$, with $C'$ being a function of the tip position over the sample.

The measurements are carried out with a sinusoidal voltage $V_{ac}$ superimposed on a dc bias $V_{dc}$, $V = V_{dc} + V_{ac} \cdot \sin(\omega t)$, where $\omega$ is angular frequency of the ac voltage and *t* is time. The amplitude of the first harmonic of the electrostatic force $F_{ac}$ can be written down as: $F_{ac} = C' \cdot V_{ac} \cdot V_{dc}$. The total electrostatic force $F_{es}$ consists of local and global components; the former is acting on the tip apex, while the latter is acting along the sensing tip cone and cantilever shank. Hence, the force $F_{es}$ can be decomposed into three contributions $F_{es} = F_{apex} + F_{cone} + F_{shank}$ as routinely done in the analysis of the electrostatic force experienced by an AFM cantilever [42, 43]. Accordingly, $C'$ can be decomposed as $C' = C'_{apex} + C'_{cone} + C'_{shank}$. Our goal will be to determine the local contribution $F_{apex}$ and its relative weight in the total force, especially in comparison with $F_{shank}$.

To probe the electrostatic forces acting on a cantilever, we performed experiments, where cantilever vibrations were excited by electrostatic forces between tip and sample as the only driving force. For that, we used a result of a previous publications [44], where it was demonstrated that the cantilever displacements in measurements on amorphous, undoped $HfO_2$ thin films deposited on a metal bottom electrode take place due to electrostatic forces alone. In particular, it was shown that charge injection from the tip apex changes the local surface potential $V_{SP}$ resulting in a hysteretic change of the AFM deflection signal at the ac frequency vs. $V_{dc}$ [44]. In this work, we used the surface potential $V_{SP}$ to drive cantilever vibrations by an ac bias voltage at $V_{dc} = 0$. In such a case, in the first harmonic, the electrostatic forces act only on the tip apex and cone with no force on the cantilever shank due to a high localization of the injected charges in the tip-sample contact area.

To obtain amplitudes of the tip apex displacement induced by the harmonic electrostatic force at the tip apex, we analyze the cantilever vibrational motion at a mechanical resonance in response to a harmonic force $F_{apex} \cdot \sin(\omega t)$ applied to the tip apex along the normal to the sample surface as shown in figure 1(a). The mathematical details of the analysis are provided elsewhere [41]. It is based on an idealized cantilever model and analytical solutions given in ref. [45]. The model is a beam of rectangular cross-section rigidly clamped at one end. Figure 1(a) displays the sensing end of the cantilever with a sensing tip. The beam is tilted in respect to the sample surface at an angle $\varphi$. A massless and rigid sensing tip is located at a distance from the beam free end and makes a right angle with the beam. Boundary conditions at the tip-sample contact are represented by two Kelvin–Voigt linear elements with springs accounting for the contact stiffness $k^*$, $k^*_{Lon}$ and dashpots describing contact damping $\gamma$, $\gamma_{Lon}$ in directions normal to and along the surface, respectively. Note that the force $F_{apex}$ acts against the normal contact stiffness $k^*$. The measurements and analysis are restricted to the cantilever vibrations in the linear regime at the first flexural contact eigenmode at resonance. The cantilever tilt angle is set to $\varphi = 12°$; cantilever dimensions are assumed to be the same as provided by manufacturers. Further, here it is assumed that $k^*_{Lon}/k^* = \gamma_{Lon}/\gamma = 0.81$ independently of the contact stiffness. To extract the apex displacement amplitudes $D_{ac}$ from the measured AFM deflection signals, static calibration of cantilever sensitivity is used with the following application of a correction factor to account for the shape of the contact resonant mode [41]. The resonant mode shapes are calculated using the analytical model of ref. [45]. The tip-sample contact stiffness $k^*$ is determined through the measured resonant frequency as described in [41].

The measurements were carried out using the band excitation (BE) technique, which is capable of measuring amplitude and phase responses of a cantilever in a frequency band encompassing a resonant peak [46, 47]. The use of such a technique is crucial here since the model outlined above requires inputs of a contact resonance frequency as well as a $Q$-factor, which cannot be extracted from single-frequency measurements. During BE, the full contact resonance peak is acquired and fitted with the single harmonic oscillator model to extract response amplitudes and resonance quality factors [46, 47].

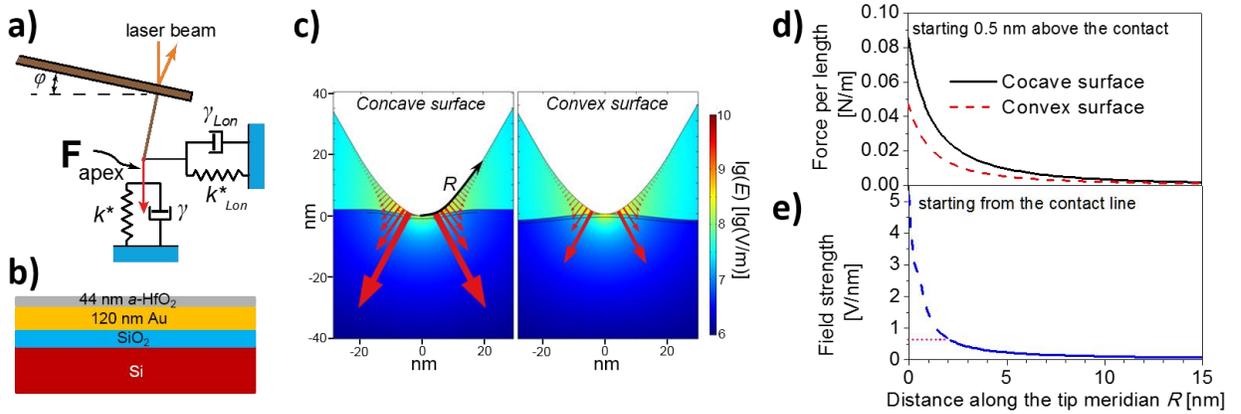

**Figure 1.** (a) Schematic of the cantilever sensing end with a sensing tip in contact with a sample. The cantilever is inclined at an angle $\varphi$ in respect to the sample surface. (b) Schematic of the sample structure used in the measurements. (c) Numerically calculated distributions of the electric field strength near the tip apex (left) and the distribution of the electrostatic force along the tip surface (right) with a voltage bias of 1 V applied to the probe for cases of convex and concave surfaces. Arrows indicate the distribution of the areal electrostatic force density (Maxwell pressure) along the tip surface. Calculations were performed with a COMSOL v.4.4 Multiphysics finite elements analysis package. Tip apex radius in the model is 20 nm and the tip-sample contact diameter is 6 nm. Permittivity of the amorphous $HfO_2$ film was set to 20. (d) Calculated electrostatic force per distance along the tip meridian (as shown with an arrow in (c), left panel) for convex and concave surfaces. (e) Calculated electric field strength along the tip meridian for a flat surface. Solid and dashed lines (blue) show the field strength calculated in the model. The dotted (pink) line indicates the field distribution near the contact as was inferred from the electrostatic force measurements.

All experiments were made with a Bruker Icon AFM in a controlled low-humidity environment (Ar-filled glove box) on a 44 nm-thick amorphous $HfO_2$ thin film sputtered on a (120 nm Au)/$SiO_2$/Si substrate. The $H_2O$ level in the glove box was <0.1 ppm as measured by the standard moisture probe in the glove box. The schematic of the sample structure is shown in figure 1(b). Such a thin-film configuration of the sample with a conducting bottom layer is highly relevant in measurements of ferroelectrics, ionic conductors, and a variety of other materials.

To shed light on relative contributions of the electrostatic forces acting on the tip apex and cone, numerical modeling was performed. Since the amorphous $HfO_2$ film has a grainy-like morphology and it surface is not perfectly smooth, it was expected that the measured signal varies with location following topography of the surface. Therefore, the modeling was done for a convex and concave surface features representing top of a grain and a grain boundary, respectively. The local surface radius for both cases was extracted from an image of the film topography. Figure 1(c) shows calculated maps of the electric field distribution around the tip apex (in logarithmic scale) when a bias of 1 V is applied to the probe. The arrows illustrate distribution of the electrostatic force density along the tip meridian. As evident from figure 1(c) and further illustrated by the curves of the force distribution along the tip meridian in figure 1(d), the force outside the tip-sample contact is strongly peaked towards the contact boundary and dissipates very quickly within 15 nm away from the contact line. Modeling shows that the 10 nm of the tip apex closest to the contact line contribute ca. 93% of the total force acting on the tip apex and cone. Therefore, the electrostatic force on the tip cone $F_{\text{cone}}$ can be neglected in the analysis. It can also be seen that the local electrostatic forces are higher for a concave surface, i.e., at a grain boundary. This is due to the fact that the distance between sample surface and tip apex is effectively smaller. The force varies with the dielectric constant $\varepsilon$ of the film material, and $\varepsilon = 20$ was set in the calculations. It should be noted that calculated forces for the convex and flat surfaces were nearly equal.

The fact that the force acting on the apex strongly dominates the overall force on the probe tip is fully consistent with the analytical result by Eliseev et al. [32] for the force acting on a sphere in contact with a dielectric. Calculations by Eliseev at al. [32] showed that the force is independent on the sphere radius, indicating a singular character of the force distribution along the sphere surface in the immediate vicinity of the contact point. Indeed, the singularity is escaped by the nature: the strong electric fields are able to activate various processes in the materials resulting in gradual transitions in the junction and reduction of the electric field strength. The exact set of processes and their relative contributions are material system-dependent and will be determined in particular by the local nanoscale surface structure, composition of adsorbates, and their behavior in high electric fields. At the same time, the high localization of the electrostatic forces in the junction results in a high resolution of the images obtained exploiting this force, which is determined by the tip-sample contact radius rather than by the long range of the electrostatic force [35, 44]. These aspects additionally justify the necessity of pixel-by-pixel measurements of the electrostatic force and determination of the coefficient $C'_{\text{apex}}$.

For a dielectric film with a surface potential $V_{\text{SP}}$, one can write:

$$F_{ac} = C' \cdot V_{ac} (V_{dc} - V_{SP}) . \tag{1}$$

From this equation, it is seen that contribution of $C'_{apex}$ in $C'$ can be calculated from the cantilever response with $V_{ac} \neq 0$, $V_{dc} = 0$, and $V_{SP} \neq 0$ locally under the apex. The local surface potential $V_{SP}$ can be readily modified by application of a large and long enough voltage pulse to the probe. Then, the first-harmonic tip displacement due to the local electrostatic force $D_{ac} = F_{apex}/k^*$ at $V_{dc} = 0$ and $V_{SP} \neq 0$ is:

$$D_{ac}(V_{SP}) = -k^{*-1} C'_{apex} V_{ac} V_{SP}. \tag{2}$$

Here, it is taken into account that the cantilever spring constant $k_C \ll k^*$, and its contribution in eq. 2 can be ignored. As seen, $D_{ac}$ changes linearly as a function of $V_{SP}$, with a slope determined by $k^*$ and $C'_{apex}$. Hence, provided that $k^*$ and $V_{SP}$ are measured independently, measurements of $D_{ac}$ open up a path to determination of the local electrostatic force and of the factor $C'_{apex}$. However, information about $V_{SP}$ as a function of $V_{dc}$ is needed. This information reflects the intensity of the charge injection at a $V_{dc}$ for a given tip-sample system and the ability of the sample surface to retain and compensate the charge for duration of measurements. The amount and retention time of the injected charge are determined by the density of the charge traps on the sample surface and in its vicinity, on the electronic energy gap and positions of the trap energy level within the electronic energy spectrum of the surface layer. Generally, $V_{SP}$ vs. $V_{dc}$ exhibits a hysteretic behavior, because occupation and liberation of available electronic traps in the surface layer and among adsorbates are associated with overcoming energy barriers by electrons [48]. Additionally, slow processes such as electric-field-induced redistribution of adsorbates on the sample surface as well as ionic motion in the sample may be involved [48, 49], which also leads to hysteresis in $V_{SP}$ vs. $V_{dc}$. The dependence $V_{SP}$ vs. $V_{dc}$ can be obtained from contact Kelvin probe force microscopy (cKPFM) measurements described in detail in ref. [44].

The cKPFM is performed with a probe in contact with a sample and involves a specific series of voltage pulses applied to the probe. After a voltage pulse of a larger amplitude, called $V_{write}$, a pulse of a smaller amplitude $V_{read}$ is applied. During application of the "writing" and "reading" pulses, a probing ac voltage $V_{ac}$ is applied on top of $V_{write}$ and $V_{read}$, and the cantilever response is detected at the frequency of $V_{ac}$. Within one measurement cycle the amplitude of $V_{write}$ is kept constant, while $V_{read}$ is varied. A typical cKPFM measurement consist of several cycles with $V_{write}$ amplitudes altered from cycle to cycle. Application of a "write" pulse, $V_{write}$, leads to charge injection into a sample. In turn, application of a series of varying "read" voltages $V_{read}$ is equivalent to application of a voltage sweep to a probe in the open-loop non-contact KPFM measurements to determine a surface potential. A "writing" pulse before each $V_{read}$ is needed to avoid the effects from a strong leakage of the injected charge back into the probe, which is in contact with the sample right at the injection site. Therefore, the series of alternating $V_{write}$ and $V_{read}$ pulses allows measuring a surface potential resulting from charge injection at a "write" voltage $V_{write}$. Performing measurement cycles with different $V_{write}$, one can obtain and measure both varying $V_{SP}$ and corresponding $D_{ac}(V_{SP})$ at $V_{dc} = 0$. Simultaneously, $V_{SP}$ as a function of $V_{write} \equiv V_{dc}$ is also obtained.

## Results and discussion

Measurements with the $HfO_2$ film were performed on a 50×50 grid in a 500 nm by 500 nm area with a probing ac bias voltage of $V_{ac} = 2$ $V_{pp}$. Figure 2(a) shows spatially averaged cKPFM curves after application of $V_{write}$ voltage pulses with varying amplitudes. Values of $D_{ac}$ were obtained from the microscope response with use of the static calibration and dynamic correction procedure outlined in the previous section. From the family of the cKPFM curves, all parameters needed to calculate $C'_{apex}$ can be extracted. Figure 2(a) displays a family of cKPFM curves $D_{ac}(V_{read})$ with $V_{write}$ being the parameter of the family. $V_{SP}$ as a function of $V_{write}$ is calculated after linear fitting each cKPFM curve $D_{ac}(V_{read})$ and determining the position of $x$-intercepts. A result of this procedure is show in figure 2(b). At the same time, from the set of data points at $V_{read} = 0$, an off-field $D_{ac}(V_{write})$ hysteresis can be extracted (figure 2(c)). The corresponding dependence of the contact stiffness vs. dc bias $k^*(V_{write})$ is displayed in figure 2(d) and shows only very small changes with the voltage. From the curves in figure 2(b)-(d), $C'_{apex}(V_{write})$ can be calculated using eq. 2. The result is shown in figure 3(a). It can be seen that $C'_{apex}$ is largely independent of $V_{write}$ (with some ill-defined points where $V_{SP}$ is close to zero). It is also of importance to note from the data in figure 1(b)-(d) that the tip apex displacement is about 2 pm/V$^2$ at a contact stiffness $k^* \approx 145$ N/m. This result may serve as a reference for estimation of the local electrostatic force contribution in microscope signal in electromechanical measurements.

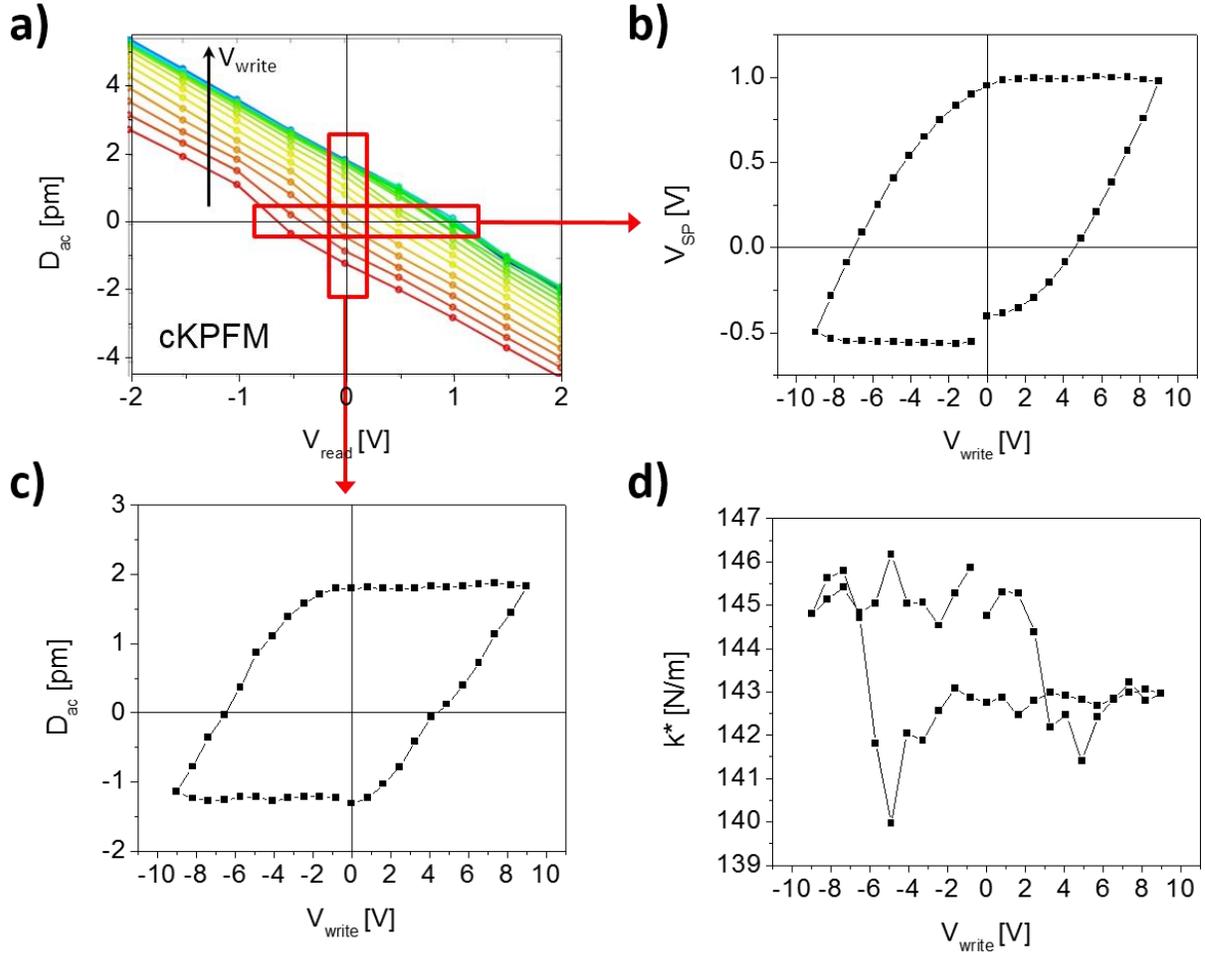

**Figure 2.** (a) cKPFM curves after application of "write" voltage pulses $V_{write}$ with varying amplitude. (b) The x-intercepts of curves in (a) are used to extract $V_{SP}$ as a function of $V_{write}$. (c) A hysteresis loop obtained for $V_{read} = 0$ from curves in (a); it is analogous to the off-field hysteresis loops in the piezoresponse force microscopy [50, 51]. (d) Contact stiffness calculated from the simultaneously measured cantilever resonance frequency. All curves are averaged over a 50×50 grid in an area of 500 nm by 500 nm and were measured with cantilever 1 in Table 1. The probing ac bias voltage was $V_{ac} = 2\ V_{pp}$

**Table 1.** Properties of the cantilevers used in the work. Length (*L*), width (*W*), thickness (*T*), free resonance frequency ($f_0$), contact resonance frequency ($f_c$), cantilever stiffness ($k_C$), contact stiffness ($k^*$), static cantilever sensitivity (*S*). *L*, *W*, and *T* are according to the manufacturers' specifications.

| # | Manufacturer | L [μm] | W [μm] | T [μm] | $f_0$ [kHz] | $f_c$ [kHz] | $k_C$ [N/m] | $k^*$ [N/m] | S [nm/V] |
|---|---|---|---|---|---|---|---|---|---|
| 1 | Budgetsensor | 225 | 28 | 2 | 73.99 | 336.7 | 3.27 | 143.3 | 116.05 |
| 2 | MikroMash | 250 | 35 | 2 | 33.6 | 171.5 | 0.48 | 64.3 | 115.99 |
| 3 | MikroMash | 250 | 35 | 2 | 33.46 | 170.0 | 0.47 | 60.5 | 126.24 |
| 4 | Nanosensor | 225 | 28 | 2 | 70 | 338.0 | 3.79 | 351.4 | 105.66 |

We further estimate relative contributions of the local and global electrostatic forces during the on-field measurements, that is, when $V_{dc}$ in eq. 1 is not zero. To do so, the obtained values of $C'_{apex}(V_{write})$ and $V_{SP}(V_{write})$ were used to analytically simulate an on-field hysteresis loop with eq. 1. The simulated loop is then compared with an experimentally measured loop acquired by applying a series of pulses of varying amplitudes $V_{dc} = V_{write}$ (with a probing ac voltage $V_{ac}$ on top of $V_{dc}$). Figure 3b shows both simulated and measured on-field loops in comparison, and it can be seen that the loops are nearly identical. This indicates that in the contact mode, the global electrostatic contribution from the cantilever shank is negligible and that only local electrostatic forces acting on the tip apex contribute to the measured signal. We, therefore, conclude that $C' \approx C'_{apex}$.

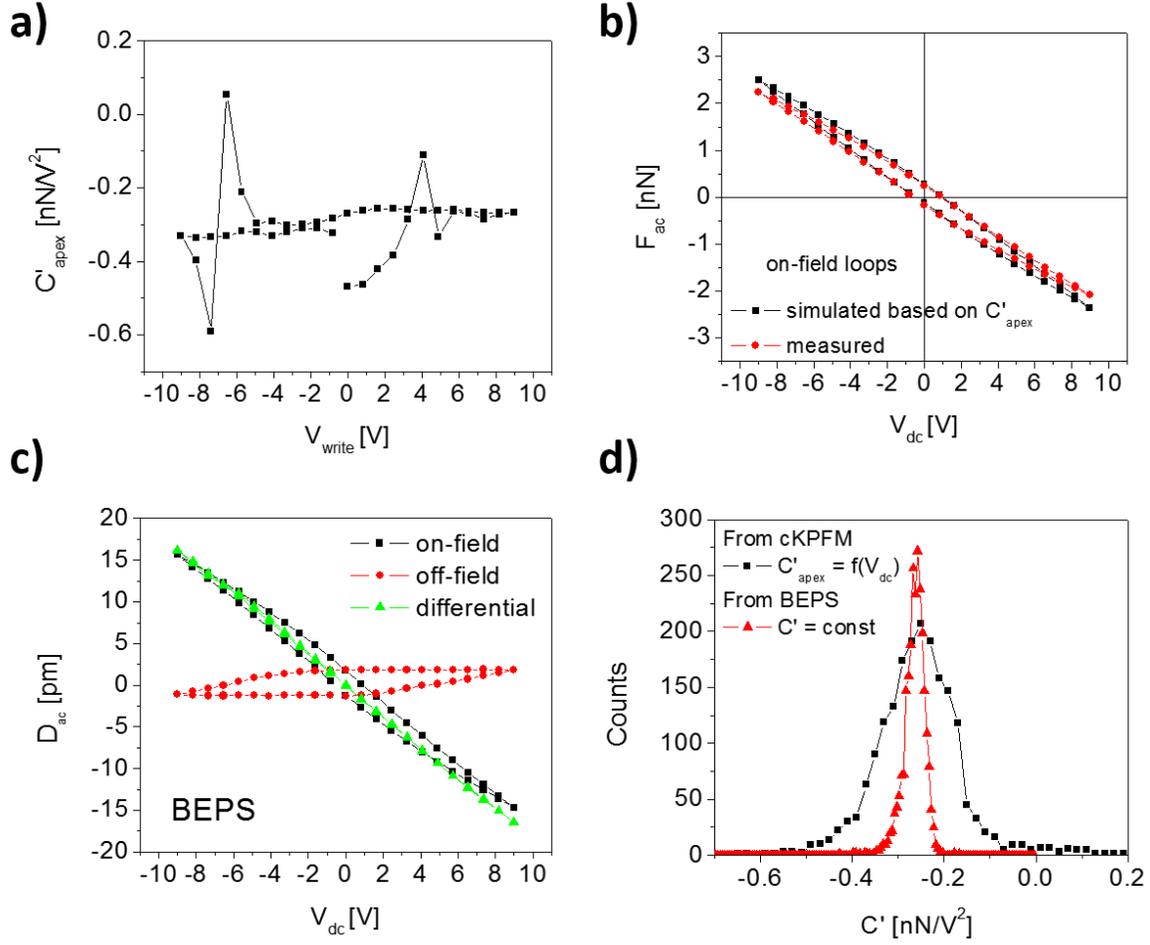

**Figure 3.** (a) Averaged $C'_{\text{apex}}$ calculated from the loops in figure 2 as a function of $V_{\text{write}}$. (b) A simulated on-field loop based on values of $C'_{\text{apex}}(V_{\text{write}})$ and $V_{\text{SP}}(V_{\text{write}})$ obtained with cKPFM measurements in comparison with a measured on-field loop. (c) Averaged on-, off-, and differential loops measured with BEPS. Curves in panels (a)-(c) are averaged over a 50×50 grid in an area of 500 nm by 500 nm. (d) Histograms of the $C'$-factor obtained from cKPFM and BEPS measurements. All measurements were performed with cantilever 1 in Table 1. The probing ac bias voltage was $V_{\text{ac}} = 2\ V_{\text{pp}}$

Owing to the fact that local electrostatic forces are dominant even during the on-field measurements, the measurement and calculation of $C'$ can be simplified. Namely, in place of the complex data acquisition and processing of cKPFM, simple measurements of hysteresis loops with Band Excitation Piezoresponse force Spectroscopy (BEPS) developed for piezoresponse force microscopy [51] can be straightforwardly used. In the BEPS voltage pulse sequence, pulses of varying amplitude $V_{dc}$ are separated by intervals with $V_{dc} = 0$. We note again that the change in surface potential is responsible for the observed hysteresis during measurements with the HfO$_2$ film leading to a noticeable off-field hysteresis (figure 2(c)). Calculating the differential loop, i.e., the difference between the BEPS on- and off-field hysteresis loops (that is, for $V_{dc} \neq 0$ and $V_{dc} = 0$, respectively), the hysteretic component can be fully removed (figure 3(c)) because the local electrostatic forces are identical in the sequential on- and off-bias states. The differential loop can be fitted to the linear equation $D_{ac} = k^{*-1} \cdot C' \cdot V_{ac} \cdot V_{dc}$ (eq. 2), where the slope is determined by $C'$ and $k^*$. As above, the contact stiffness $k^*$ is obtained from the cantilever resonance frequency during the data acquisition. However, in the simplified procedure, $C'$ is calculated from the slope of the $D_{ac}(V_{dc})$ line using the known $k^*$. Such analysis was performed for every point of the 50×50 grid, and the corresponding histogram is shown in figure 3(d) in comparison with the data obtained with the cKPFM-based method using the $V_{SP}(V_{write})$ dependence and off-field hysteresis loops. Both approaches yield the same mean value of $C'$. This once more supports the conclusion that in the contact mode, the local electrostatic forces dominate over the global ones. As seen in figure 3(d), the differential loops method yields a narrower distribution of $C'$. This is due to the higher fitting accuracy of the method, since the data are fit in a wider voltage window of $V_{dc}$ as compared to $V_{SP}$, yielding a larger signal-to-noise ratio.

Several metal-coated probes with spring constants ranging between 0.47 N/m and 3.79 N/m were used to measure the factors $C'$ using the BEPS-based method. Probe properties are provided in Table 1. As illustrated in figure 4(a), each cantilever model shows a different $C'$. However, if the same cantilever model is used, the values $C'$ are close (cantilevers 2 and 3). This is most likely due to the high sensitivity of the electrostatic forces to the particular tip apex shape, which varies between probes of different manufacturers. To demonstrate the spatial variability of $C'$, figure 4(b) shows a map of $C'$ measured with cantilever 4 in Table 1. As seen in the map, the force on the tip apex is higher at grain boundaries and the microstructure is easily revealed by the map. Interpretation of this fact is directly follows from the results of finite elements modeling shown in figure 1(c)-(e).

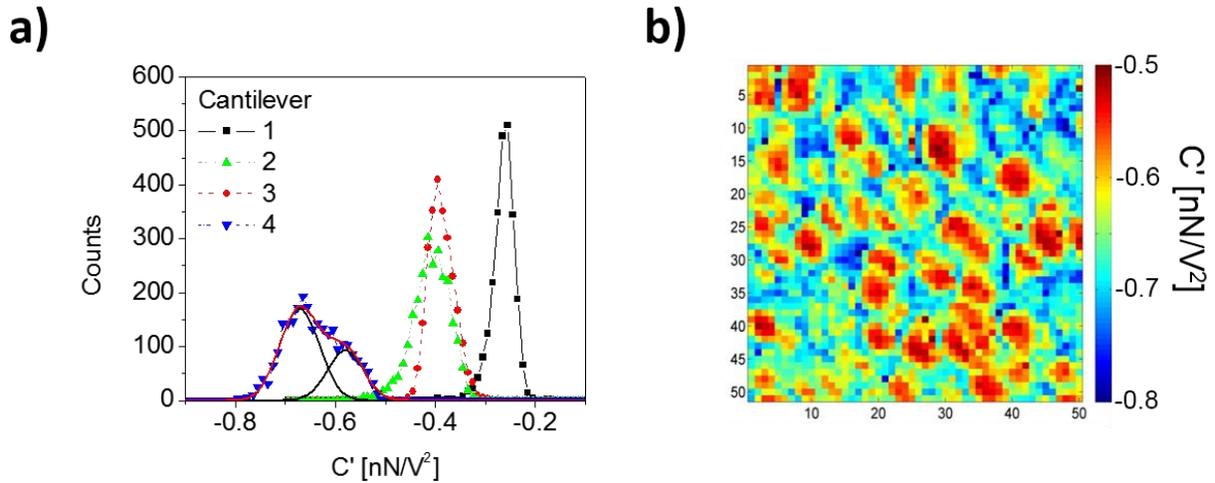

**Figure 4.** (a) Histogram of factor $C'$ obtained from BEPS measurements performed with different cantilevers. Cantilever properties are provided in Table 1. Cantilevers 2 and 3 are of the same model and have slightly different characteristics. A two-peaks fit for the histogram for cantilever 4 is also shown. (b) A map of local variation of the factor $C'$ acquired with cantilever 4 in Table 1.

It is of great interest to compare the experimentally obtained values of $C'$ with values of the numerical simulations. In the simulations, the total force on the tip apex and cone is calculated by integrating the force distributions like those shown in figure 1(d) over the length along the tip meridian. The numerical modeling yields $F = 1.13$ nN for the convex surface model shown in figure 1(c) at a tip bias $V = 1$ V. Since $F = \frac{1}{2} C' \cdot V^2$, the corresponding value of $C'$ is 2.26 nN/V². What is immediately striking from this result is the much larger—from 3 to 10 times—values of this factor in comparison with the experimental ones. We also note the extremely large values of the maximal electric field in the tip-sample junction of the model (figure 1(e)), which reaches $5\times10^9$ V/m. Apparently, the large force in the model is due to these extremely large values of the field, which may not be present in the real tip-sample junction. The field strength in the model should be reduced to about $0.6\times10^9$ V/m as illustrated in figure 1(e) to results in $C' \approx 0.5$ nN/V$^2$, which is in the range of the experimentally obtained values. In fact, this indirectly provides an estimate of the maximal electric field strength in the tip-sample junction. As discussed before by other authors [9-11], and evidenced by measurements of this work, this field is 1 V/nm in the order of magnitude. However, the necessity to reduce the field in comparison with the calculated value, strongly suggests that the field is close to the maximum, which can be achieved in the tip-sample junction. Besides the charge injection and other processes mentioned in Introduction, multiple processes induced by such a strong field may include motion of contaminants on the sample and tip apex surfaces (caused by Maxwell stress) as well as ionic and electronic transport processes in the sample surface layer and in the junction vicinity. It is worth noting specifically that a field strength of 0.5 V/m creates a voltage drop of about 0.2 V across a distance of 0.4 nm, which is close to a typical unit cell size in simple oxides and other compounds or to a typical jump distance determining intensity of charge transport processes in solids. For many materials activation energies of such processes are below or close to 1 eV, and therefore, their kinetics is expected to be significantly intensified in the vicinity of the probe-sample contact. All these can effectively lead to reduction of the field strength. More detailed measurements of the electrostatic force for different sample structures and materials as well as analysis of the possible processes induced by the high electric field in the tip-sample junction are planned in the near future. However, it should be noted that the sample configuration studied here—a thin dielectric film and a conducting bottom

electrode—is one of the most frequently encountered in electromechanical SPM experiments, and therefore, the results presented above can be widely generalized.

## Conclusions

In summary, contact-resonance-enhanced AFM measurements was used to quantify the local electrostatic forces on the probe tip apex due to a non-zero surface potential as well as when an electrical bias is applied to a probe in experiments with a non-ferroelectric amorphous $HfO_2$ thin film. Comparing the results, we can conclude that global electrostatic forces, which scale with dimensions of the cantilever, are negligible in the contact mode measurements on a thin film with a bottom electrode. We show by modeling and in experiment that the local topography has a direct effect on the electrostatic forces acting on the tip apex and that the force strongly depends on details of the tip apex shape. By comparing modeling and experiment, the maximal electric field strength in the tip-sample junction was estimated to be about 0.5 V/nm. Such a strong field may lead to large alterations of multiple charge and mass transport processes in the junction vicinity as well as to shifts of phase equilibria. Besides a method to measure local electrostatic force and the electric field strength in the tip-sample junction, our results provide a baseline and a way to account for the effects of local electrostatic forces in electromechanical AFM measurements as well as offer additional means to probe ionic mobility and field-induced phenomena in solids.

## Acknowledgements

Experiments were planned and conducted through personal support provided by the U.S. Department of Energy, Basic Energy Sciences, Materials Sciences and Engineering Division through the Office of Science Early Career Research Program (N.B.). The facilities to perform the experiments were provided at the Center for Nanophase Materials Sciences, which is sponsored at Oak Ridge National Laboratory by the Scientific User Facilities Division, Office of Basic Energy Sciences, U.S. Department of Energy, which also provided additional personal support (S. J., B. C., M. B. O., I. K., S.V.K., A.T.). A.T. also acknowledges CICECO-Aveiro Institute of Materials (Ref. FCT UID/CTM/50011/2013) financed by national funds through the FCT/MEC and, when applicable, co-financed by FEDER under the PT2020 Partnership Agreement. I.K. provided the HfO$_2$ sample.